\newcommand{\be}{\begin{equation}}\newcommand{\ee}{\end{equation}}
\newcommand{\bea}{\begin{eqnarray}}\newcommand{\eea}{\end{eqnarray}}
\newcommand{\nn}{\nonumber}\newcommand{\p}[1]{(\ref{#1})}
\newcommand{\lb}{\label}

\newcommand{\cA}{{\cal A}}
\newcommand{\bcA}{\bar{\cal A}}

\newcommand\T{\mbox{Tr}\;}

\newcommand\q{\quad}
\newcommand\qq{\qquad}

\newcommand{\pp}{{=\!\!\!|}}
\newcommand{\xp}{x^\pp}
\newcommand{\xm}{x^=}

 \newcommand{\Pp}{\partial_\pp}
\newcommand{\Pm}{\partial_=}

\newcommand{\Pxi}{\partial_\xi}
\newcommand{\bPxi}{\bar\partial_\xi}

\newcommand{\tpi}{\theta^+_i}

\newcommand{\tmi}{\theta^-_i}

\newcommand{\btpi}{\bar{\theta}^{i+}}

\newcommand{\btmi}{\bar{\theta}^{i-}}

\newcommand{\bppt}{\bar{\partial}_{2+}}

\newcommand{\bpmt}{\bar{\partial}_{2-}}
\newcommand{\bpmh}{\bar{\partial}_{3-}}
\newcommand{\bpph}{\bar{\partial}_{3+}}
\newcommand{\ppo}{\partial^1_+}
\newcommand{\ppt}{\partial^2_+}
\newcommand{\pmo}{\partial^1_-}
\newcommand{\pmt}{\partial^2_-}

\newcommand{\Dpk}{D_+^k}

\newcommand{\bDpk}{\bar{D}_{k+}}

\newcommand{\tpo}{\theta^+_1}
\newcommand{\tpt}{\theta^+_2}
\newcommand{\tmo}{\theta^-_1}
\newcommand{\tmt}{\theta^-_2}
\newcommand{\tmh}{\theta^-_3}
\newcommand{\tph}{\theta^+_3}
\newcommand{\btpo}{\bar\theta^{1+}}
\newcommand{\btmo}{\bar\theta^{1-}}
\newcommand{\btpt}{\bar\theta^{2+}}
\newcommand{\btmt}{\bar\theta^{2-}}
\newcommand{\btph}{\bar\theta^{3+}}
\newcommand{\btmh}{\bar\theta^{3-}}
\newcommand{\Dpo}{D_+^1}
\newcommand{\Dmo}{D_-^1}
\newcommand{\Dpt}{D_+^2}
\newcommand{\Dmt}{D_-^2}

\newcommand{\bDph}{\bar{D}_{3+}}
\newcommand{\bDmh}{\bar{D}_{3-}}

\newcommand{\Dot}{D^1_2}
\newcommand{\Dto}{D^2_1}
\newcommand{\Doh}{D^1_3}

\newcommand{\Dth}{D^2_3}

\newcommand{\hDot}{\hat{D}^1_2}

\newcommand{\hDth}{\hat{D}^2_3}

\documentstyle[12pt]{article}

\begin{document}
\begin{flushright}
{ hep-th/0008148 }
\end{flushright}
\vspace{2cm}

\begin{center}
{\large\bf
HARMONIC-SUPERSPACE TRANSFORM FOR  N=3 SYM-EQUATIONS }
\vspace{0.5cm}

{\bf Ji\v{r}\'{\i} Niederle~$^a$ and Boris Zupnik~$^b$}\\

{\it $^a$Institute of Physics, Academy of Sciences of the Czech Republic,
Prague 8, CZ 182 21, Czech Republic}\\
{\it$^b$ Bogoliubov Laboratory of Theoretical Physics, Joint Institute
for Nuclear Research, Dubna, Moscow Region, 141980, Russia}\\
\end{center}
\begin{abstract}
 The $SU(3)/U(1)\times U(1)$ harmonic variables are used in  the
harmonic-superspace representation of the $D{=}4,~N{=}3$ SYM-equations.
The  harmonic superfield equations of motion in the simple non-covariant
gauge contain the nilpotent harmonic analytic  connections. It is shown
that these harmonic SYM-equations are equivalent to the finite set of
solvable linear iterative equations.
\end{abstract}
\renewcommand{\thefootnote}{\arabic{footnote}}
\setcounter{footnote}0
\setcounter{equation}0
\section{Introduction}

The $SU(2)/U(1)$ harmonic superspace (HSS) has been used
to solve the $D=4, N=2$ off-shell constraints \cite{GIK1,Zu2}.
The integrability interpretation of the dual version of the equations
of motion has been also formulated for
 the $N=2$ supersymmetric Yang-Mills (SYM) theory in the framework
of this superspace \cite{Zu4}.

In the HSS-approach to the $D=4,\;N=3$ supersymmetry ,
the $SU(3)/U(1)\times U(1)$ harmonics have been used for the off-shell
description of the corresponding SYM-theory \cite{GIKOS}.
The $N=3$ SYM-equations in the ordinary superspace \cite{So} have been
transformed to the zero-curvature equations for the harmonic
gauge connections, however, nobody tried earlier to solve these equations.
Moreover, it has been shown that some set of
analytic connections does not generate any solutions in the ordinary
superspace \cite{RS}.

We propose to  solve first the harmonic zero-curvature equations in terms
of the independent non-analytic superfield  matrix $v$ ( bridge)
describing the transform between different representations of the
gauge group.  Then we analyze the  dynamical Grassmann (G-)
analyticity equations for the composed on-shell harmonic connections
constructed in terms of this superfield $v$.

 It will be shown that  the special non-covariant gauge choice for the
matrix $v$ simplifies drastically the solution of all equations.
The crucial feature of our gauge is the nilpotency of $v$ and the
corresponding G-analytic harmonic connections. The Lorenz invariance of
the SYM-equations is broken down to the $SO(1,1)$ subgroup in this gauge.
We demonstrate that the 1-st order harmonic bridge equations with
the nilpotent HSS-connections produce also the  linear 2-nd
order differential constraints for the bridge matrix.

Thus, we show that the $N=3$ SYM-equations in the ordinary superspace
can be transformed to the linear solvable matrix differential equations
in the harmonic superspace. This method allow us to analyze also
solutions of the dimensionally reduced SYM-equations with 12 supercharges.

\setcounter{equation}0
\section{\lb{B}Harmonic representation of $D=4,~N=3$ SYM
 constraints }

We shall consider  the non-covariant
$SO(1,1)\times SU(3)$
representation of the $D=4, N=3$ superspace coordinates $z$
\bea
&&x^{1\dot{1}}\equiv \xp=t+x^3~,\q x^{2\dot{2}}\equiv \xm=t-x^3~,\q
x^{1\dot{2}}\equiv y=x^1+ix^2~,\q
 x^{2\dot{1}}\equiv \bar{y}=x^1-ix^2~,\nn\\&&
\theta^1_i=\tpi~,\q\theta^2_i=\tmi~,\q \bar\theta^{i\dot{1}}=\btpi~,\q
\bar\theta^{i\dot{2}}=\btmi
~.\lb{A2}
\eea
where $1, 2,~\dot{1}, \dot{2}$ are the $SL(2,C)$ indices,
 $i=1, 2$ and 3 are indices of the fundamental
representations of the  group $SU(3)$.
The $SO(1,1)$ weights of these coordinates are $0, \pm1, \pm2$,
respectively.

Let us introduce the $(4|6,6)$-dimensional superspace gauge connections
$A(z)$ and the corresponding covariant derivatives $\nabla$
\bea
&&\nabla^i_\pm=D^i_\pm + A^i_\pm~,\qq\bar{\nabla}_{i\pm}=\bar{D}_{i\pm} +
\bar{A}_{i\pm}~,\lb{A7}\\
&&\nabla_\pp =\Pp + A_\pp ~,\qq\nabla_= =\Pm + A_= ~,\q
\nabla_y=\partial_y+A_y~,\q \bar\nabla_y=\bar\partial_y+\bar{A}_y~,\nn
\eea
where the space-time derivatives and the Grassmann derivatives are
considered.

 The $D=4,~N=3$ SYM-constraints \cite{So} have the following
reduced-symmetry form:
\bea
&&\{\nabla^k_+,\nabla^l_+\}=0~,\q \{\bar{\nabla}_{k+},
\bar{\nabla}_{l+}\}=0~,
\q \{\nabla^k_+,\bar{\nabla}_{l+}\}=2i\delta^k_l\nabla_\pp~,\lb{A8}\\
&&\{\nabla^k_+,\nabla^l_-\}=\bar{W}^{kl}
~,\q\{\nabla^k_+,\bar{\nabla}_{l-}\}=2i\delta^k_l \nabla_y~,\lb{A9}\\
&&\{\nabla^k_-,\bar{\nabla}_{l+}\}=2i\delta^k_l\bar{\nabla}_y~,\q
\{\bar{\nabla}_{k+},\bar{\nabla}_{l-}\}=W_{kl}~,\lb{A10}\\
&&\{\nabla^k_-,\nabla^l_-\}=0~,\q \{\bar{\nabla}_{k-},
\bar{\nabla}_{l-}\}=0~,
\q \{\nabla^k_-,\bar{\nabla}_{l-}\}=2i\delta^k_l\nabla_=~,\lb{A11}
\eea
where $W^{kl}$ and $\bar{W}^{kl}$ are the gauge-covariant
superfields constructed from the gauge connections.
The equations of motion for the superfield strengthes follow from the
Bianchi identities.

Let us analyze first Eqs.\p{A8} which can be treated as integrability 
conditions for the positive-helicity connections.
These conditions have the following pure gauge solution:
\be
\nabla^k_+=g^{-1}\Dpk g~,\q \bar{\nabla}_{k+}=g^{-1}\bDpk g~,\q
\nabla_\pp =g^{-1}\Pp g~.\lb{A13}
\ee

Using the  on-shell gauge condition $g=1$ we can obtain the simple
general solution of
Eqs.\p{A8}
\be
A^k_+=0~,\q \bar{A}_{k+}=0~,\q A_\pp =0~.\lb{A14}
\ee

The analogous gauge conditions excluding the part of
connections has been considered in Ref.\cite{DL} for the self-dual
$4D$ SYM-theory and in Ref.\cite{GS} for the 10D SYM equations.

The $SU(3)/U(1)\times U(1)$ harmonics \cite{GIKOS} parameterize
the corresponding coset space. They form an $SU(3)$ matrix $u^I_i$ and
are defined modulo $U(1)\times U(1)$ transformations
\be
u^1_i=u^{(1,0)}_i\;,\q u^2_i=u^{(-1,1)}_i\;,\q
u^3_i=u^{(0,-1)}_i\;,\lb{A15}
\ee
where $i$ is the index of the triplet representation of $SU(3)$. The
complex conjugated harmonics have opposite $U(1)$ charges
\be
u^i_1=u^{i(-1,0)}~,\q u^i_2=u^{i(1,-1)}\;,\q u^i_3=u^{i(0,1)}\;.
\lb{A16}
\ee

These harmonics satisfy the following relations:
\bea
&& u_i^I u^i_J=\delta^I_J\;,\q u^I_i u^k_I=\delta^k_i\;,\nn\\
&&\varepsilon^{ikl}u_i^1 u_k^2 u_l^3=1\;.\lb{A17}
\eea

The $SU(3)$-invariant harmonic derivatives act on the harmonics
\bea
&&\partial^I_J u^K_i =\delta^K_J u^I_i\;,\q \partial^I_J u^i_K=-
\delta^I_K u^i_J\;,\nn\\
&&[\partial^I_J,\partial^K_L]=\delta^K_J\partial^I_L-\delta^I_L
\partial^K_J\;.\lb{A18}
\eea

We shall use the special $SU(3)$-covariant conjugation
\be
\widetilde{u^1_i}=u^i_3\;,\q \widetilde{u^3_i}=u^1_i\;,\q
\widetilde{u^2_i}=-u^2_i\;.\lb{A19}
\ee
The corresponding conjugation of the harmonic derivatives is
 \be
\widetilde{\partial^1_3 f}=-\partial^1_3\widetilde{f}~,\qq
\widetilde{\partial^1_2 f}=\partial^2_3\widetilde{f}~,
\lb{A20}
\ee
where  $f(u)$ is a harmonic function.

We can define the real analytic harmonic superspace $H(4,6|4,4)$
with 6 coset harmonic dimensions and the following left and right
coordinates:
\bea
&&\zeta=(\xi^\pp~,~\xi^=~,~\xi~,~\bar{\xi}~,~ \theta^\pm_2~,
~\theta^\pm_3~,~\bar\theta^{1\pm}~,~\bar\theta^{2\pm})~,\nn\\&&
\xi^\pp=\xp +i(\tph\btph -\tpo\btpo)\;,\qq\xi^= =\xm +
i(\tmh\btmh -\tmo\btmo)\;,\nn\\
&& \xi=y+i(\tph\btmh -\tpo\btmo)\;,\qq\bar\xi=\bar{y}
+i(\tmh\btph -\tmo\btpo)\;,\lb{A22}
\eea
where
$\theta^\pm_I=\theta^\pm_k u^k_I~,\q\bar\theta^{\pm I}=
\bar\theta^{\pm k}u_k^I$.

The  CR-structure in $H(4,6|4,4)$ involves the G-derivatives
\be
D^1_\pm,\;\bar{D}_{3\pm}~,
\lb{A23}
\ee
which commute with the harmonic derivatives $\Dot,\;\Dth$ and $\Doh$.

These derivatives have the following explicit form in the
analytic coordinates:
\bea
&&D^{(1,0)}_\alpha\equiv D^1_\alpha=\partial^1_\alpha\equiv
\partial/\partial\theta^\alpha_1\;\q
\bar{D}^{(0,1)}_\alpha\equiv \bar{D}_{3\alpha}=\partial_{3\alpha}\equiv
\partial/\partial \bar\theta^{3\alpha}\;,\lb{A24}\\
&&D^{(2,-1)}\equiv \Dot =\partial^1_2
+{i\over2}\tpt\btpo\Pp+{i\over2}\tpt\btmo\Pxi+{i\over2}\tmt\btpo\bPxi
+{i\over2}\tmt\btmo\Pm\nn\\
&&-\tpt\ppo-\tmt\pmo+\btpo\bppt+\btmo\bpmt
~,\nn\\
&&D^{(-1,2)}\equiv\Dth =\partial^2_3
+{i\over2}\tph\btpt\Pp+{i\over2}\tph\btmt\Pxi+{i\over2}\tmh\btpt\bPxi
+{i\over2}\tmh\btmt\Pm\nn\\
&&-\tph\ppt-\tmh\pmt+\btpt\bpph+\btmt\bpmh\;,\lb{F9}\\
&&D^{(1,1)}\equiv\Doh =\partial^1_3
+i\tph\btpo\Pp+i\tph\btmo\Pxi+i\tmh\btpo\bPxi
+i\tmh\btmo\Pm\nn\\
&&-\tph\ppo-\tmh\pmo+\btpo\bpph+\btmo\bpmh\;,\lb{A25}
\eea
where $\Pp =\partial/\partial \xi^\pp,~\Pm =\partial/
\partial \xi^=,~\Pxi=\partial/\partial\xi$ and $\bPxi=
\partial/\partial\bar\xi$.

It is crucial that we start from  the specific  gauge conditions
\p{A14} for the $N=3$ SYM-connections which break  $SL(2,C)$ but 
preserve the $SU(3)$-invariance. Consider the harmonic transform
of the  covariant Grassmann derivatives in this gauge using the 
projections on the $SU(3)$-harmonics
\bea
&&\nabla^I_+\equiv u_i^I D^i_+=D^I_+\;,\q
\bar\nabla_{I+}\equiv u^i_I\bar{D}_{i+}=\bar{D}_{I+}\;,\q
\{D^I_+,\bar{D}_{K+}\}=2i\delta^I_K\Pp\;,\lb{D1b}\\
&&\nabla^I_-\equiv u_i^I\nabla^i_-=D^I_- +\cA^I_-\;,\q
\bar\nabla_{I-}\equiv u^i_I\nabla_{i-}=\bar{D}_{I-} +\bcA_{I-}\;,
\lb{D1}
\eea
where the harmonized Grassmann connections $\cA^I_-$ and $\bcA_{I-}$
are defined.

The $SU(3)$-harmonic projections of the superfield constraints
(\ref{A9}-\ref{A11}) can be derived from the basic set of the
$N=3$ super-integrability conditions for two components
of the harmonized connection:
\bea
&&\Dpo\cA^1_-=\bDph\cA^1_-=\Dpo\bcA_{3-}=\bDph\bcA_{3-}=0\;\lb{D2}\\
&&\Dmo\cA^1_- +(\cA^1_-)^2=0\;,\q \bDmh\bcA_{3-}+(\bcA_{3-})^2=0\;,
\nn\\
&&\Dmo\bcA_{3-}+\bDmh\cA^1_- +\{\cA^1_-,\bcA_{3-}\}=0\;.\lb{D3}
    \eea
All projections of the SYM-equations can be obtained by the action
of the harmonic $SU(3)$ derivatives $D^I_K$ on these basic conditions.

These Grassmann zero-curvature equations have the very simple general
solution
\be
\cA^1_-(v)=e^{-v}\Dmo e^v\;,\q \bcA_{3-}(v)=e^{-v}\bDmh e^v\;,\lb{D4}
\ee
where {\it the bridge} $v$ is the gauge-Lie-algebra valued 
(5,5)-superfield matrix satisfying the additional  constraint
\be
(\Dpo,\;\bDph) v=0\;.\lb{D5}
\ee

Consider the gauge transformations of the bridge
\be
e^v\;\Rightarrow\;e^\lambda e^v e^{\tau_r}\;,\lb{D11}
\ee
where $\lambda\in H(4,6|4,4)$ is the (4,4)-analytic matrix parameter,
and the parameter $\tau_r$ \lb{A14b} does not depend on harmonics.
The matrix $e^v$ realizes the harmonic transform of the gauge superfields
$ A^k_\pm, \bar{A}_{k\pm}$ in the central basis (CB) to the harmonic
gauge superfields of the analytic basis (AB) using the analytic
$\lambda$-representation of the gauge group.

The  bridge $v$ defines the on-shell harmonic connections of the
analytic basic
\be
V^I_K(v)=e^vD^I_K e^{-v}\lb{D10}
\ee
which satisfy by construction the harmonic
zero-curvature equations, for instance,
\bea
&& \Dot V^2_3-\Dth V^1_2 +[V^1_2,V^2_3]-V^1_3=0\;,\nn\\
&&\Dot V^1_3-\Doh V^1_2 +[V^1_2,V^1_3]=0\;.\lb{hzcr}
\eea

The dynamical SYM-equations in the bridge representation \p{D4}
are reduced to the  harmonic differential conditions for
the basic Grassmann connections:
\be
(D^1_2,\;D^2_3,\;D^1_3)\left(\cA^1_-(v),\;\bcA_{3-}(v)\right)=0\;,
\lb{Hanal}
\ee
which are equivalent to the following set of the dynamic G-analyticity
equations:
\be
(\Dmo,~\bDmh) \left( V^1_2(v), V^2_3(v), V^1_3(v)\right)=0\;.\lb{D7}
\ee

The positive-helicity analyticity conditions
\be
 (\Dpo,~\bDph)\left( V^1_2(v), V^2_3(v), V^1_3(v)\right)=0\lb{D12}
\ee
are satisfied automatically for the bridge in the gauge  \p{D5}.
Stress that the analyticity equations in the bridge representation
describe all SYM-solutions without a loss of generality.

The inverse harmonic transform  determines the harmonic-independent
on-shell gauge solutions in the ordinary superspace
\bea
&& \cA^1_-(v)~\Rightarrow~A^i_-(v)=(u^i_1+u^i_2D^2_1+u^i_3D^3_1)
e^{-v}\Dmo e^v\;\nn\\
&&\bcA_{3-}(v)~\Rightarrow~\bar{A}_{i-}(v)=(u_i^3-u^1_iD^3_1-u^2_iD^3_2)
e^{-v}\bDmh e^v\;.
\lb{K8}
\eea
By construction, these superfields satisfy the $D=4,~N=3$ CB-constraints
(\ref{A9}-\ref{A11}) and the harmonic differential conditions
\be
D^I_K\left(A^i_-(v),\bar{A}_{i-}(v)\right)=0~,
\ee
if the relations \p{Hanal} are fulfilled.

It is useful to calculate the Grassmann connection $\cA^1_-(v)$
in terms of the basic analytic matrices
\be
e^{-v}\Dmo e^v=b^1-\tmo (b^1)^2+\btmh \left(d^1_3-{1\over2}\{b^1,c_3\}
\right)+
\tmo\btmh\left([b^1,d^1_3]+{1\over2}[c_3,(b^1)^2]\right)\;.\lb{K9}
\ee
The conjugated connection $\bcA_{3-}(v)$ can be constructed analogously.

\setcounter{equation}0
\section{Harmonic-superspace equations of motion}

Using the $\lambda$-transformations \p{D11} one can gauge away
the first terms in the Grassmann decomposition of the
bridge $v$ and choose the non-supersymmetric nilpotent  gauge for
this (5,5)-superfield:
\bea
&&v=\tmo b^1+ \btmh c_3 +\tmo\btmh d^1_3\;,\qq v^2=\tmo\btmh
[c_3,b^1]\;,\lb{D12c}\\
&&e^{-v}= I - v+{1\over2}v^2=I-\tmo b^1- \btmh c_3+\tmo\btmh
({1\over2}[c_3,b^1]-d^1_3)\;,
\lb{D12b}
\eea
where the fermionic   matrices $b^1, c_3 $ and the bosonic matrix
$d^1_3 $ are (4,4)-analytic functions of the coordinates $\zeta$ \p{A22}.

The restrictions on the bridge matrix in the gauge group $SU(n)$
\be
\T v=0\;,\qq v^\dagger=-v\lb{D15}
\ee
 are equivalent to the following relations
for the analytic matrices  $b^1, c_3 $ and $d^1_3 $
\be
\T b^1=0\;,\q \T c_3=0\;,\q \T d^1_3=0\;
,\q(b^1)^\dagger=c_3\;,\q (d^1_3)^\dagger=-d^1_3
\;.\lb{D14}
\ee

It is useful to construct the on-shell superfield strength in the
analytic basis
\be
\bar{W}^{12}(b^1)
=-\Dpt b^1+\tmt\Dmt\Dpt b^1
+\tmt[\Dto b^1,\Dpt b^1]
+\tmo\tmt[\Dmt b^1,\Dpt b^1]\;,\lb{H6}
\ee
which satisfies the non-Abelian conditions of Grassmann and
harmonic analyticities.

The dynamical G-analyticity equation \p{D7} in the gauge \p{D12c}
is equivalent to the following harmonic differential bridge equation:
\bea
&&V^1_2(v)=\tmt b^1\equiv v^1_2\;,\qq (v^1_2)^2=0\;\lb{dyn}\\
&&\Dot e^{-v}=e^{-v}v^1_2~.\lb{br}
\eea
where the  analytic nilpotent representation for the on-shell harmonic
connection arises.

The 2-nd on-shell harmonic connection is also nilpotent
\bea
&&V^2_3(v)=-\btmt c_3\equiv v^2_3\;,\lb{prep2}\\
&&\Dth e^{-v}=e^{-v}v^2_3\;.\lb{br2}
\eea

 Underline that the nilpotency of the analytic parts of
the harmonic connections $V^1_2$ and $V^2_3$ \p{D10} follows directly
 from the gauge condition \p{D12}, and the analyticity of
these connections requires the additional harmonic restrictions on the
functions $b^1,c_3$ and $d^1_3$.

The 1-st bridge equation \p{br} generates the following nonlinear 
equations for the (4,4)-analytic matrices:
\bea
&& \Dot b^1=-\tmt (b^1)^2\;,\lb{D17}\\
&&\Dot c_3=-\tmt(d^1_3+{1\over2}\{b^1,c_3\})\;,
\lb{D18}\\
&&\Dot d^1_3={1\over2}\tmt\left([d^1_3,b^1]+{1\over2}[(b^1)^2,c_3]
\right)~.\lb{D19b}
\eea

It is important that all these equations contain the nilpotent
element $\tmt$ in the nonlinear parts, so they can be
reduced to the set of linear iterative equations.

The 2-nd bridge equation \p{br2} gives us
\bea
&&\Dth b^1=\btmt(-d^1_3+{1\over2}\{b^1,c_3\})\;,
\lb{D20}\\
&&\Dth c_3=\btmt (c_3)^2\;\lb{D21}\\
&&\Dth d^1_3={1\over2}\btmt\left([d^1_3,c_3]+{1\over2}[b^1,(c_3)^2]
\right)~.\lb{D22}
\eea

It is useful to derive the following relations:
\bea
&&\tmt\Dot(b^1,~c_3,~d^1_3)=0~,\q \btmt\Dth ( b^1,~c_3,~d^1_3)=0~,
\lb{D23}\\
&&\tmt\Dth b^1+\btmt\Dot c_3=\tmt\btmt\{b^1,c_3\}~,\lb{D24}\\
&&\btmt\Dot c_3-\tmt\Dth b^1=2\tmt\btmt d^1_3~,\lb{D25}\\
&&\tmt\Doh b^1+\btmo\Dot c_3=-\tmt\tmh(b^1)^2+\tmt\btmo\{b^1,c_3\}~.
\eea

The bridge equations with the nilpotent analytic connections  yield
the following 2-nd order linear conditions for the coefficient functions
\be
\Dot\Dot (b^1, c_3,  d^1_3)=0\;,\q \Dth\Dth (b^1, c_3,  d^1_3)
=0\;.\lb{lin2}
\ee

Now we consider the iterative procedure of solving
the basic non-Abelian harmonic differential equations for the
(4,4) analytic matrices $b^1$ and $c_3$ using the partial decomposition
in the Grassmann variables $\tmt, \tmh, \btmo, \btmt$.
The matrix (4,0) coefficients of this decomposition have
dimensions $-1/2 \ge l \ge -5/2$
\bea
&&b^1=\beta^1+\tmt B^{12}+\tmh B^{13}+\btmo B^0+\btmt B^1_2
+\tmt\tmh \beta^0+ \tmt\btmo \beta^2+\tmh\btmt \beta^{13}_2\nn\\
&&+
\tmh\btmo \beta^3+\btmo\btmt \beta_2
+\tmt\btmt\eta^1+\tmt\tmh\btmo B^{23}
+\tmh\btmo\btmt  B_2^3+\tmt\tmh\btmt C^{13}\nn\\
&&+\tmt\btmo\btmt C^0+\tmt\tmh\btmo\btmt\eta^3~.\lb{I2}
\eea

The first iterative (4,0) equations
$(l=-1/2)$ are linear and homogeneous
\be
\hDot\beta^1= \hDth \beta^1=0~.\lb{I13}
\ee

The next harmonic  iterative equations for the  (4,0) components
with $l \le -1/2$ can be  resolved on the each step via  functions of
the highest dimensions or their  derivatives.
These linear equations contain  nonlinear sources
constructed from the solutions of the previous  equations.
Note that some (4,0) iterative equations are pure algebraic relations
which reduce the number of independent functions, for instance,
\bea
&& B^1_2=-\hDot B^0-{i\over2}\tpt\Pxi\beta^1~,\q
B^{12}=\hDth B^{13}+{i\over2}\btpt\bPxi\beta^1~,\lb{I5}\\
&&\beta^2=\hDth \beta^3-{i\over2}\btpt\bPxi B^0~,\q
\beta^{13}_2=-\hDot \beta^3+{i\over2}\tpt\Pxi B^{13}~.\lb{I4}
\eea

The finite set of the linear harmonic differential
equations for the independent (4,0) functions $\beta^1, B^0, B^{13},
\beta^3\ldots$ can be, in principle, explicitly solved. Thus, the
$SU(3)/U(1)\times U(1)$ harmonic method together with the simple
gauge conditions allow us to transform the $N=3$ superfield
SYM-constraints  to the harmonic differential equations  which are
equivalent to the finite set of the iterative solvable linear  equations.

The authors are grateful to E.A. Ivanov for the discussions.

This work is supported by the Votruba-Blokhintsev programme
 in Joint Institute for Nuclear Research.
The work of B.Z. is partially supported also by
the grants RFBR-99-02-18417, RFBR-DFG-99-02-04022
and NATO-PST.CLG-974874.

\end{document}